\documentclass[aps,showpacs]{revtex4}

\usepackage{amssymb,amsthm}
\newcommand{\R}{{\mathbb R}}

\newcommand\be{\begin{eqnarray}}
\newcommand\ee{\end{eqnarray}}

\begin{document}

\title{Plebanski gravity without the simplicity constraints}

\author{Kirill Krasnov}
\affiliation{School of Mathematical Sciences, University of Nottingham, NG7 2RD, UK}
  
\date{November 2008}

\begin{abstract} In Plebanski's self-dual formulation general relativity becomes SO(3)
BF theory supplemented with the so-called simplicity (or metricity) constraints for the
B-field. The main dynamical equation of the theory states that the curvature of the B-compatible SO(3) connection
is self-dual, with the notion of self-duality being defined by the B-field. We describe a theory
obtained by dropping the metricity constraints, keeping only the requirement that the curvature of the
B-compatible connection is self-dual. It turns out that the theory one obtains is to a very large degree
fixed by the Bianchi identities. Moreover, it is still a gravity theory, with 
just two propagating degrees of freedom as in GR. 
\end{abstract}


\maketitle

\section{Introduction}

Vacuum Einstein equations can be elegantly obtained as follows. Consider the Riemann curvature tensor 
$R_{\mu\nu\rho}^{\quad\,\sigma}$ of a metric $g_{\mu\nu}$ on a 4-manifold $M$. A natural way to obtain a second
order differential equations for $g$ is to require that the curvature is ``proportional'' to the metric. 
However, since the curvature has many more components than the metric, it is natural to consider
the Ricci curvature $R_{\mu\nu}=R_{\mu\rho\nu}^{\quad\,\rho}$, which is, as the metric itself, a
symmetric tensor. One can now impose the condition $R_{\mu\nu}=\lambda g_{\mu\nu}$, where $\lambda$ is, to start with, 
an arbitrary function on $M$. Making use of the Bianchi identity $\nabla^\nu G_{\mu\nu}=0$, where
$G_{\mu\nu}=R_{\mu\nu}-(1/2)R g_{\mu\nu}$, one finds that $\lambda$ must be a constant, referred to by 
physicists as (a multiple of) the cosmological constant.

In 1977 Plebanski \cite{Plebanski} proposed an elegant reformulation of general relativity that
is based not on symmetric rank two tensors (metrics), but on anti-symmetric rank two ones (two-forms).
The main idea uses the notion of self-duality
on two-forms and is as follows. It is well-known that one can rewrite the vacuum Einstein equations
$R_{\mu\nu}\sim g_{\mu\nu}$ as a condition stating that the the self-dual part 
of the full Riemann curvature tensor, where the self-duality is taken with respect to second pair of indices, 
is also self-dual with respect to the first pair: 
\be\label{einst}
P^- R P^+ =0.
\ee
Here $P^\pm =(1/2)({\rm Id}\pm (1/i)*)$ are the self- and anti-self dual projectors, 
and the star in the second term denotes 
the Hodge operator $A^*_{\mu\nu}:=(1/2)\epsilon_{\mu\nu}^{\quad\rho\sigma} A_{\rho\sigma}$ on two-forms,
with a mixed tensor $\epsilon_{\mu\nu}^{\quad\rho\sigma}$ obtained from the volume 4-form
$\epsilon_{\mu\nu\rho\sigma}$ of $g_{\mu\nu}$ by raising two of its indices with the metric.

Plebanski theory reformulates general relativity in a way that directly leads to
(\ref{einst}). To this end one introduces a (complexified in Lorentzian signature) SO(3) 
vector bundle $V$ over $M$, which can be
referred to as the self-dual bundle, and a two-form field $B^i, i=1,2,3$ taking values in $V$.
The triple of two forms $B^i, i=1,2,3$ encodes information about the metric $g$ on $M$ via the requirement 
that $B^i, i=1,2,3$ are self-dual two-forms with respect to $g$. Indeed, the triple $B^i$ spans a
3-dimensional subspace in the space of all two-forms, and declaring this to be the
subspace of self-dual two forms defines the notion of Hodge duality on two forms, which, in turn, 
can be shown \cite{Samuel} to uniquely determine the conformal class of the metric. A representative 
in this conformal class is then fixed by requiring that $(i/3)(B^i\wedge B^i)$ is the volume form of $g$. 
However, a general triple $B^i$ of two-forms contains too many components. Indeed, it needs $3\times 6$ 
numbers to be specified, while a metric has only 10 components. To remedy this, Plebanski imposes
the following ``metricity'' (or simplicity) conditions:
\be\label{metr}
B^i \wedge B^j \sim \delta^{ij},
\ee
which give 5 equations on the two-form field (the trace of this equation gives the proportionality
coefficient and is an identity). This brings the number of components
in $B^i$ down to 13, which is the required 10 components describing the metric, plus
3 gauge components related to availability of SO(3) gauge transformations. 

Thus, supplemented by the metricity conditions (\ref{metr}) the two-form field contains
just the right amount of information to describe a metric. One now needs a second order
differential equation on $B^i$. To obtain this one notices that there is a unique
connection $A^i$ satisfying:
\be\label{comp}
D_A B^i =0,
\ee
where $D_A B^i:=dB^i + \epsilon^{ijk} A^j\wedge B^k$.
Indeed, this gives $4\times 3$ algebraic equations for $4\times 3$ components of $A^i$,
which fixes it uniquely, provided a certain non-degeneracy conditions for $B^i$ are
satisfied. We shall refer to this connection as $B$-compatible. When $B^i$ satisfies
the metricity conditions (\ref{metr}) the $B$-compatible connection turns out to
be equal to the self-dual part of the metric-compatible one. One can now compute the curvature 
$F^i:=dA^i + (1/2)\epsilon^{ijk}A^j\wedge A^k$ of the $B$-compatible connection.
A natural second order equation on $B^i$ is then obtained by requiring that the
curvature $F^i$ is ``proportional'' to the two-form field $B^i$. However, we will
now allow for the ``proportionality coefficient'' to be an ``internal'' tensor:
\be\label{eq-Pleb}
F^i = \Phi^{ij} B^j,
\ee 
where the quantities $\Phi^{ij}$ are at this stage arbitrary. We note that the equation
(\ref{eq-Pleb}) can be considered an analog of Einstein's condition $R_{\mu\nu}\sim g_{\mu\nu}$,
but now in the context of Lie algebra-valued rank two anti-symmetric tensors (two-forms) instead
of rank two symmetric tensors. It also captures the content of equation (\ref{einst}), since it gives 
precisely the requirement that the self-dual part of the curvature (equal to the curvature of the 
self-dual connection $A^i$), is self-dual as a two-form.

As in the case of the Einstein condition $R_{\mu\nu}\sim g_{\mu\nu}$ the coefficients
$\Phi^{ij}$ introduced above can be further restricted by means of Bianchi identities. Thus, one can 
show that the trace part $\Lambda=-{\rm Tr}(\Phi)$ must be a constant (the cosmological constant),
while the trace-free part $\Psi^{ij}$ is arbitrary (symmetric). The dynamical equations
(\ref{eq-Pleb}) are then 18 equations for 13 unknown functions contained in $B^i$ plus
5 unknown functions contained in $\Psi^{ij}$. As (\ref{einst}) shows, equations 
(\ref{eq-Pleb}) are just Einstein equations in disguise, so we obtain a reformulation
of vacuum general relativity in which the metric never appears directly. Importantly, one can
also convert into the two-form framework the right-hand-side of Einstein equations -
the stress-energy momentum tensor, but we shall not consider the non-vacuum case
in this short paper.

An action that leads to all the equations above is given by:
\be\label{action-Pleb}
S[B,A,\Psi]=\int B^i\wedge F^i(A) - \frac{1}{2}\left( \Psi^{ij} - \frac{\Lambda}{3}\delta^{ij}
\right) B^i\wedge B^j.
\ee
Indeed, the variation with respect to the traceless tensor $\Psi^{ij}$ gives (\ref{metr}), 
variation with respect to the connection gives (\ref{comp}), while variation with
respect to the two-form field gives the main dynamical equation (\ref{eq-Pleb}).

The aim of this short paper is to analyze Plebanski's theory described above with the
metricity conditions (\ref{metr}) removed. Our main result is that the theory so
obtained is to a large extent fixed by the Bianchi identities and is 
specified by one arbitrary function $\Lambda(\Psi)$ of the field $\Psi^{ij}$. Most interestingly,
for any (generic) choice of this function the theory contains just two propagating degrees of freedom
as in GR. The class of theories with varying $\Lambda(\Psi)$ has been introduced in an
earlier work \cite{Krasnov:2006du} of the author from very different (renormalization) considerations.
It is also worth noting that the same class of theories has 
been described in much earlier works by Bengtsson and Peldan under the name of 
``neighbours of GR'', see \cite{Bengtsson:1990qg} for their first appearance. These authors'
starting point was the so-called pure connection formulation, so the equivalence to 
the theory described in \cite{Krasnov:2006du} was not immediately obvious and was pointed out
in \cite{Bengtsson:2007zzd}. The results of this paper provide a different, more geometrical 
perspective on this interesting class of gravity theories. 

As is clear from (\ref{action-Pleb}), in Plebanski formulation GR becomes a theory of BF-type,
in that its Lagrangian starts with the same term as the BF theory one, and the field content
is similar. We shall see that the new gravity theory obtained by dropping the simplicity
constraints is also of the BF type. Moreover, as we shall see, it can be cast into a theory even closer
to BF by eliminating the $\Psi$ field. The theories of BF type fall into the category
quantizable using the so-called spin foam model techniques. These work by discretizing
the theory and then attempting to ``deform'' the known simplicial quantization of BF
theory to produce a theory that is close to (or, ideally, coincides with) a theory of
interest. These ideas have recently led to some important progress, see
\cite{Engle:2007uq} and \cite{Freidel:2007py} in the field of spin foam models
of quantum gravity. A description of gravity given in this paper suggests a new
take on the idea of spin foam quantization, as we shall comment on in the last section.

The paper is organized as follows. In Section \ref{sec:removed} we analyze the
Plebanski theory with simplicity constraints removed and show how this theory
is almost uniquely fixed by the Bianchi identities. We conclude this short paper 
with a discussion.

\section{Simplicity constraints removed}
\label{sec:removed}

As we have already described in the Introduction, in this paper we would like to consider
a theory similar to Plebanski's gravity in that its main dynamical fields are a Lie-algebra valued
two-form field $B^i$ and a connection $A^i$. The idea is to write down a consistent
system of second-order differential equations for the two-form field $B^i$. Recalling that
any configuration of the two-form field gives rise to a spacetime metric (by declaring
the two-forms $B^i$ to be self-dual with respect to this metric and choosing a volume form), we will 
thus get a gravity theory with second-order field equations.

We have already discussed in the Introduction that given a two-form field $B^i$ (not necessarily
satisfying any simplicity constraints, but being non-degenerate in a certain precise sense), there is 
a unique $B$-compatible connection $A_B$ that satisfies $D_{A_B} B^i = 0$. An explicit formula
for this connection is available in e.g. \cite{Bengtsson:1995rz}, but we will not need it in this paper.
Having the $B$-compatible connection $A_B$ one can compute its curvature $F(A_B)$. It is
given by some complicated expression involving up to second derivatives of $B^i$. The
curvature is a two-form, and it can be decomposed into the basis of self-dual 
$B^i$ and anti-self-dual $\bar{B}^i$ two-forms:
\be\label{F-decomp}
F^i(A_B) = M^{ij} B^j + N^{ij} \bar{B}^j.
\ee
The anti-self-dual forms satisfy:
\be\label{reality}
B^i \wedge \bar{B}^j = 0,
\ee
and can in principle be determined once $B^i$ are known. In the case of Lorentzian
signature general relativity one further requires the anti-self-dual two-forms
to coincide with the complex conjugates of the two-forms $B^i$: $\bar{B}^i = (B^i)^*$.
In this case (\ref{reality}) become the reality conditions for the conformal
metric determined by $B^i$. 

So far this is completely general and no field equations are imposed. 
Let us now write down the field equations. For these we shall keep the same equations
as in the Plebanski case (\ref{eq-Pleb}), and require the curvature of the
connection $A_B$ to be purely self-dual:
\be\label{field-eqs-vac}
F^i(A_B)=\Phi^{ij} B^j \qquad \Longleftrightarrow \qquad M^{ij}=\Phi^{ij},
\quad N^{ij} = 0,
\ee
where $\Phi^{ij}$ is some purely gravitational tensor to be described below.
The system of equations (\ref{field-eqs-vac}) gives 18 equations for the
18 components of the two-form field $B^i$. However, it also contains the
so-far unspecified functions $\Phi^{ij}$ and so is not complete. Note that
in the GR case we have exactly the same system of 18 equations, but in that
case for 18-5 quantities $B^i$ (the two-form field $B^i$ modulo the conditions
$B^i\wedge B^j\sim\delta^{ij}$). In addition the trace part of $\Phi^{ij}$
is either zero (no cosmological constant case) or constant (cosmological 
constant), and is thus not an unknown field. The system of 18 equations
is thus that for 13 components of $B^i$ and the remaining 5 components
of $\Phi^{ij}$.

In the general case the system of equations (\ref{field-eqs-vac}) can be completed 
by considering the analogs of ``Bianchi'' identities.
Thus, we note that the components of $M^{ij}, N^{ij}$ in (\ref{F-decomp}) are
not independent. Indeed, we have the following Bianchi identity:
\be
D_{A_B} F(A_B) = 0.
\ee
This gives:
\be\label{bianchi-1}
(D_{A_B} M^{ij}) \wedge B^j + (D_{A_B} N^{ij} \bar{B}^j)=0.
\ee
Another important identity is obtained by using the compatibility equations
$D_{A_B} B^i=0$. Taking another covariant derivative and using the
definition of the curvature we get:
\be\label{bianchi-2}
\epsilon^{ijk} F^j(A_B) \wedge B^k = 0 \Longleftrightarrow \epsilon^{ijk} M^{jl} B^l\wedge B^k =0.
\ee
This last equation can be conveniently interpreted as follows. Let us define a conformal ``internal'' metric:
$B^i\wedge B^j \sim h^{ij}$. Then (\ref{bianchi-2}) can be rewritten as:
\be
\epsilon^{ijk} M^{jl} h^{lk} =0.
\ee

Let us now also introduce an action principle that leads to (\ref{field-eqs-vac}) as
Euler-Lagrange equations. This is easy to write, we have:
\be\label{action-full-vac}
S[B,A,\Phi]=\int B^i\wedge F^i(A) - \frac{1}{2}\Phi^{ij} B^i\wedge B^j.
\ee
Varying this with respect to $A^i$ we get $D_A B^i=0$, which requires $A$ to be the
$B$-compatible one, varying the action with respect to $B^i$ we get (\ref{field-eqs-vac}).
We also note that only the symmetric part of the field $\Phi^{ij}$ enters the
action, so it is necessary to assume that $\Phi^{ij}$ in (\ref{field-eqs-vac}) is symmetric
if we are to have an action principle for our theory.

It remains to clarify the meaning of the variation with respect to $\Phi^{ij}$. To
these end we shall use the Bianchi identities (\ref{bianchi-1}), (\ref{bianchi-2}).
Using (\ref{bianchi-1}) and field equations (\ref{field-eqs-vac}) we see that we
must have:
\be
D_{A_B} \Phi^{ij}\wedge B^j =0.
\ee
Let us multiply this equation by the one-form $\iota_\xi B^i$ and sum over $i$. Here
$\xi$ is an arbitrary vector field and  $\iota_\xi B^i$ is one-form with components
$(\iota_\xi B^i)_\mu:=\xi^\alpha B_{\alpha\mu}^i$. However, for any vector field $\xi$ we have:
\be\label{key-ident}
\iota_\xi B^{(i} \wedge B^{j)} = \frac{1}{2} \iota_\xi (B^i \wedge B^j) \sim h^{ij},
\ee
where $h^{ij}$ is the internal metric introduced above. This gives us the following equation:
\be
h^{ij} D_{A_B} \Phi^{ij} = 0.
\ee
Now, using the symmetry of $h^{ij}$ we can rewrite this equation as 
$h^{ij}(d\Phi^{ij} + 2\epsilon^{ikl} A^k \Phi^{lj})=0$. However,
the other Bianchi identity (\ref{bianchi-2}) together with the field equation
$M^{ij}=\Phi^{ij}$ implies $\epsilon^{kli} \Phi^{lj} h^{ji} =0$ and so we must have:
\be\label{h-Phi-ident}
h^{ij} d\Phi^{ij}=0.
\ee

The identity (\ref{h-Phi-ident}) implies that the quantities $h^{ij}$ and $\Phi^{ij}$ are
not independent. This can be seen quite clearly by considering the last term in 
the action (\ref{action-full-vac}). Using the introduced above tensor $h^{ij}$ we
can write the integrand as $V:=h^{ij} \Phi^{ij}$ times some volume form. We now have:
\be\label{dV}
dV=\Phi^{ij} dh^{ij} + h^{ij} d\Phi^{ij} =  \Phi^{ij} dh^{ij},
\ee
where we have used (\ref{h-Phi-ident}). This means that (i) the last term
in the action is only a function of the $h^{ij}$ components of the two-form
field $B^i$; (ii) the quantities $\Phi^{ij}$ are also expressible through
$h^{ij}$ and are given by:
\be\label{Phi-h}
\Phi^{ij} = \frac{\partial V(h^{ij})}{\partial h^{ij}}.
\ee
Below we shall characterize the ``potential'' $V(h^{ij})$ in more details.
For now let us note that having expressed the unknown functions $\Phi^{ij}$ in
terms of the the components of the two-form field $B^i$ we have closed the
system of equations (\ref{field-eqs-vac}), as it is now a system of 18 equations
for 18 unknowns - components of the $B^i$ field.

It is also instructive to see what the derived identity (\ref{h-Phi-ident}) boils
down to in the case of GR. In that case $h^{ij}\sim\delta^{ij}$ and so we have
$d{\rm Tr}(\Phi)=0$, which implies that the trace part of the field $\Phi^{ij}$ must
be a constant - the cosmological constant. Thus, the identity (\ref{h-Phi-ident}) is
a generalization of this well-known in GR requirement, obtained in exactly the
same way as a consequence of field equations and Bianchi identities. 

It is now convenient to rewrite (\ref{h-Phi-ident}) in the following manner.
Thus, let us decompose both $h^{ij}$ and $\Phi^{ij}$ into their trace and traceless parts:
\be
h^{ij} \sim \delta^{ij} + H^{ij}, \qquad \Phi^{ij} = \Psi^{ij} - \frac{\Lambda}{3}\delta^{ij},
\ee
where $H^{ij}, \Psi^{ij}$ are both traceless, and we have written the trace part of
$\Phi^{ij}$ in a way suggestive of the cosmological constant, which we know this part
is in the case of GR. Of course in our more general case $\Lambda$ is so-far an arbitrary
function of spacetime coordinates. The identity (\ref{h-Phi-ident}) becomes:
\be\label{h-Psi-ident}
H^{ij} d\Psi^{ij} = d\Lambda,
\ee
and the above introduced potential becomes:
\be
V = H^{ij} \Psi^{ij} - \Lambda.
\ee
One easily recognizes in these relations those of a Legendre transform between two
functions. Indeed, we have:
\be
dV = \Psi^{ij} dH^{ij} + H^{ij} d\Psi^{ij} - d\Lambda = \Psi^{ij} dH^{ij},
\ee
where we have used (\ref{h-Psi-ident}). This means that the potential $V=V(H^{ij})$
is the Legendre transform of the function 
\be
\Lambda = \Lambda(\Psi),
\ee
and that
\be\label{eqs-Psi}
H^{ij}=\frac{\partial \Lambda(\Psi^{ij})}{\partial \Psi^{ij}}.
\ee
This last relation arises now directly from the action, as the field equation
obtained when varying with respect to $\Psi^{ij}$. Indeed, 
the action (\ref{action-full-vac}) becomes:
\be\label{action-full}
S[B,A,\Psi]=\int B^i\wedge F^i(A) - \frac{1}{2}\left( \Psi^{ij} - \frac{\Lambda(\Psi)}{3}\delta^{ij}
\right) B^i\wedge B^j.
\ee
The Euler-Lagrange equations following by varying this action with respect to
$\Psi^{ij}$ are exactly (\ref{eqs-Psi}). It is also clear that the field equations 
(\ref{eqs-Psi}) imply the identity (\ref{h-Psi-ident}), which we have derived
by considering Bianchi identities. This closes the set of equations of the
theory in the sense that all field equations become consistent with the
identities between them, as well as in the sense that the main field equations
(\ref{field-eqs-vac}) become a set of 18 equations for 18 unknowns $B^i$. 

At least classically, an equivalent way to describe the theory (\ref{action-full}) is 
in terms of the ``potential'' $V(h^{ij})$ introduced above. In (\ref{Phi-h}) we have seen that 
the Bianchi identities relate the fields $\Phi^{ij}$ to partial derivatives 
of the potential $V(h^{ij})$ with respect to the components of the ``internal''
metric $h^{ij}$. We can then rewrite the action (\ref{action-full}) in terms
of only the fields $B^i, A^i$ as:
\be\label{action-pot}
S[B,A]=\int B^i\wedge F^i(A) - \frac{1}{2} V(h_B^{ij}) (vol)_B, 
\ee
where we have defined the internal metric $h_B^{ij}$ as a function of the $B$-field via: 
\be\label{h}
B^i \wedge B^j = h_B^{ij} (vol)_B, \qquad (vol)_B = \frac{1}{3} B^i\wedge B^i
\ee
so that ${\rm Tr}(h)=3$. Note that we have introduced the volume form $(vol)_B$ as
defined by the $B$-field. The action (\ref{action-pot}) has the form of BF
theory with a ``potential'' for the components $h_B^{ij}$ of the $B$-field that are
extracted via (\ref{h}). The potential can be arbitrary. The most
striking fact about (\ref{action-pot}) is that it is still a gravity 
theory in the sense that it is a theory of metrics that propagates just two degrees 
of freedom, as in GR. We shall provide some further explanation for how this
can be possible in the next section.

To summarize, we have seen that the condition $B^i\wedge B^j\sim \delta^{ij}$ of Plebanski
formulation of GR can be relaxed and how the Bianchi identities still lead
(in a unique way) to a consistent theory. Note that what one obtains is  
a class of gravity theories rather than one theory, for a theory is now specified by 
a choice of function $\Lambda(\Psi)$ (or $V(h)$), which can be completely arbitrary. The
nature of the ``modification'' as compared to the GR case can be summarized by
saying that in the new theories the cosmological function has become a 
function of the ``curvature'' $\Psi^{ij}$. 

\section{Discussion}

There are several remarks that should be made about the gravity theory (\ref{action-full}),
(\ref{action-pot}). First, what we have described above can be thought of as an embedding
of Einstein's gravity theory (in its Plebanski version) into a much larger class of
theories. In this embedding a certain constraint term
that was present in the Plebanski action was replaced by a potential term. There is
an illuminating quantum field theory textbook example in which a similar embedding
occurs. Namely, let us consider a nonlinear sigma model whose dynamical field is $n^i \in \R^N$
that is constrained to lie on the unit sphere. The corresponding Lagrangian is:
\be\label{non-lin}
{\mathcal L} = \frac{1}{2g^2} (\partial_\mu n^i)^2  - \psi((n^i)^2 - 1).
\ee
Here $g$ is the coupling constant of the theory and the last term contains a Lagrange multiplier $\psi$ 
that imposes the sphere condition $n^i n^i =1$. This theory describes $N-1$ massless bosons, but is 
non-renormalizable in dimensions higher than two, as follows e.g. from the fact that the mass
dimension $[g]=(2-d)/2$ of its coupling constant is negative for $d>2$. 

The above nonlinear sigma model can be embedded into a simpler (and renormalizable) theory by replacing
the constraint in (\ref{non-lin}) with a potential. Indeed, consider the following Lagrangian for 
a field $\phi^i\in \R^N$:
\be\label{linear}
{\mathcal L} = \frac{1}{2} (\partial_\mu \phi^i)^2 + \frac{1}{2} \mu^2 (\phi^i)^2 - \frac{\lambda}{4} [(\phi^i)^2]^2. 
\ee
The potential above is minimized for any $\phi_0^i$ that satisfies $(\phi_0^i)^2=\mu^2/\lambda$. Choosing
a vacuum breaks the ${\rm O}(N)$ symmetry down to ${\rm O}(N-1)$, and the
spectrum of excitations around the vacuum chosen is that of $N-1$ massless Goldstone bosons and
one massive mode (of mass $2\mu^2$). Making the potential infinitely steep (in the direction
in which it has a non-vanishing second derivative) we send the mass of the massive mode to
infinity and produce the theory of $N-1$ massless bosons - the above described nonlinear
sigma-model. The massive mode present in (\ref{linear}) and absent in (\ref{non-lin})
is what in the standard model is known as the Higgs particle.

What happens in the passage from Plebanski theory (\ref{action-Pleb}) to (\ref{action-pot})
is similar, but with one crucial difference: no additional degree of freedom is added in
(\ref{action-pot}) as compared to (\ref{action-Pleb}). In this sense, one can refer to the mechanism 
described in this paper as ``Higgs without Higgs''. Let us see how this happens in more details.

Introducing the ``internal metric'' $h_B^{ij}$ constructed out of the $B$-field as
in (\ref{h}), we can put the Plebanski theory action (\ref{action-Pleb}) into a form similar to that
of nonlinear sigma-model:
\be\label{action-constr}
S[B,A]=\int_M B^i\wedge F^i - \frac{1}{2} \Psi^{ij}\left( h_B^{ij} - \delta^{ij} \right) (vol)_B,
\ee
where $(vol)_B$ is the volume form defined by $B$, see (\ref{h}). This action is exactly
of the nonlinear sigma-model form ``kinetic term + constraint''. Indeed, after one solves
for the connection in terms of the derivatives of the $B$-field and substitutes the solution
into the first term, one gets a term of the form $(dB)^2$, which is just a kinetic term for
$B$. The second, constraint term, puts the $B$-field on 
the surface $h_B^{ij}=\delta^{ij}$, which is analogous to the sphere $n^i n^i =1$ in nonlinear
sigma-model. After this is done, the kinetic $BF$ term gives an analog of 
the term $(\partial_\mu n^i)^2$.

Changing to the theory (\ref{action-pot}), we see that the only difference as compared
to (\ref{action-constr}) is that the constraint $h_B^{ij}=\delta^{ij}$ was
replaced by a potential for $h_B^{ij}-\delta^{ij}$. The kinetic term is unchanged.
Essentially, we see the structure of the action (\ref{linear}) with all the
components of the $B$-field being present and a potential for the fluctuations
of the components $h_B^{ij}$ around $\delta^{ij}$ added. Unlike the case
(\ref{non-lin}), (\ref{linear}), there is now not one but five additional modes $h_B^{ij}$ added
to the original Plebanski theory. But the key difference is that the modes
that have been added do not propagate. What makes this possible is that the
kinetic term $(dB)^2$ that arises from the $BF$ term of the original action is degenerate. 

The degeneracy of the $BF$ term is very well-known, and is at the heart of
the topological invariance of this theory. A detailed account of why  
no additional degrees of freedom are present in (\ref{action-pot})
as compared to (\ref{action-Pleb}) is given in \cite{Krasnov:2007cq}. Here
we would like to present a simplified version of this story. Thus,
let us consider a simple classical mechanics
example that mimics the mechanism at work in the passage from (\ref{action-Pleb})
to (\ref{action-pot}). For this, let us consider a totally constrained system with momentum 
and position variables $p,q$ and the symmetry $p\to p + \alpha$ that is described by the
following action: $S= \int dt (p\dot{q} - \lambda q)$. Here $\lambda$ is a
Lagrange multiplier that sets $q=0$ and generates shifts of the momentum
variable. The system described is an over-simplified version of $BF$ theory,
with $\lambda$ being the analog of $B_{0a}^i$ components of the $B$-field,
the later being the BF theory Lagrange multipliers - generators of
its topological symmetry. 

Let us now change the system by adding
to it an additional Lagrange multiplier $\psi$: $S= \int dt (p\dot{q} - \lambda q - \psi \lambda)$.
The effect of the Lagrange multiplier $\psi$ is to set the original Lagrange multiplier,
generator of the symmetry $p\to p+\alpha$ to zero and thus introduce degrees
of freedom. The resulting system is that describing a particle with
zero Hamiltonian. This is essentially the mechanism of how the
degrees of freedom appear in Plebanski formulation of GR. The main
difference between the presented oversimplification and the real
story is that not all of the $B_{0a}^i$ Lagrange multipliers of BF theory are set
to zero. Four of them remain, and the constraints they are associated
to generate the spacetime diffeomorphisms.

To describe an analog of (\ref{action-pot}), 
let us now, instead of adding a constraint setting $\lambda=0$, add
a potential term for $\lambda$. Thus, consider the following action:
$S= \int dt (p\dot{q} - \lambda q - V(\lambda))$. Now $\lambda$ is
no longer a Lagrange multiplier, and it is best to introduce the
momentum $\pi_\lambda$ conjugate to it. One gets $\pi_\lambda\approx 0$
as a constraint. Commuting this constraint with the Hamiltonian one gets
$q + V_{,\lambda}\approx 0$. This is a secondary constraint that, together
with $\pi_\lambda\approx 0$ forms a second class system whenever the
second derivative of the potential $V_{,\lambda\lambda}$ is non-zero. For
a generic potential this is so almost everywhere, so one gets two second class
constraints. One then solves the $q$-constraint for $\lambda$ and substitutes the solution
back into the action. One gets a particle with the Hamiltonian given
by the Legendre transform of the potential function $V(\lambda)$. The
simple story presented exactly mimics the Hamiltonian analysis \cite{Krasnov:2007cq}
of the modified gravity theories (\ref{action-pot}) and provides an illustration for why no additional
degrees of freedom are introduced as compared to Plebanski theory. Thus, we see that the key point
for why it was possible to go from a nonlinear sigma-model-like Plebanski 
theory to a linear sigma-model-like one (\ref{action-pot}) is that
the kinetic term of the action (\ref{action-pot}) is degenerate. Moreover,
it is degenerate precisely in those directions which are being set to zero by the 
Lagrange multipliers in Plebanski theory and appear in the potential
in the modified theory. 

The sigma-model analogy described, seems especially relevant because the non-renormalizability of 
the nonlinear sigma model, as well as its resolution in the linear model, is often compared
to the non-renormalizability of quantum gravity. Indeed, the nonlinear
model is non-renormalizable because one ``forgot'' about an additional degree of
freedom - the Higgs boson. Once this is added into the picture, the obtained
theory becomes renormalizable. Thus, the non-renormalizability of the
nonlinear sigma-model is rightfully interpreted as signalling an additional
degree of freedom that will start playing role at higher energies. It
is widely believed that the same reasoning is applicable to quantum gravity,
and that its non-renormalizability signals that additional degrees of freedom
should become important at high energies and make the theory renormalizable. 

However, we have just seen an embedding of a non-renormalizable nonlinear sigma-model-like
Plebanski theory into a linear sigma-model-like theory (\ref{action-pot}). While
we do not yet know whether the theory (\ref{action-pot}) is renormalizable, we
saw that no new degrees of freedom were added.
Even the very fact that this is possible is at first surprising. And this fact suggests
that maybe the sought UV completion of the gravity theory may be not so drastically
different from the low energy theory - GR. Indeed, the very existence of the
class of theories (\ref{action-pot}) suggests that may be one should look
for this UV completion in a class of theories that, as GR, have just two
propagating degrees of freedom. This intuition would be even more supported
should it be found that the class of theories (\ref{action-pot}) is renormalizable.
Work is currently in progress to see whether this may be the case. But whatever
the renormalizability properties of the theories (\ref{action-pot}) are, their
very existence provides an interesting counterargument to the seemingly irrefutable ``new degrees
of freedom in quantum gravity'' intuition.

Our second remark concerns the so-called spin foam approach to quantum gravity. 
This is based on the understood simplicial quantization of BF theory, with 
the main idea being to modify the BF state sum to produce a 
model for quantum gravity. Correspondingly, most of the effort in the field of spin 
foam quantization of gravity goes into the question of how the simplicity constraints on the 
$B$-field can be consistently imposed at the discrete level, see \cite{Engle:2007uq} and
\cite{Freidel:2007py} for recent progress in this direction. As no discrete 
way of imposing the {\it self-dual} Plebanski constraints is known, in the spin foam approach 
one currently works with the ``full'' Plebanski theory without the chiral split. In this
paper, motivated by the above construction, we would like to propose
a possibility of a rather drastic alternative to this paradigm. Namely, as we have 
seen, dropping the simplicity constraints does not lead to any
serious change in the nature of the theory. One can keep oneself as close
to general relativity as one wants by keeping the potential steep enough. Thus, 
we propose that it could be fruitful to change the viewpoint from ``gravity=BF theory + constraints'' 
to ``gravity = BF theory + potential''. The benefit of such a reformulation is
immediate: one certainly has more intuition about dealing with potentials than with constraints
in quantization. This would also allow one to work with a simpler (at least in the
Euclidean signature where there are no reality conditions to impose) self-dual theory.
It remains to be seen, however, whether such a change in the point of view
can lead to any progress in quantum gravity.

\bigskip
{\bf Acknowledgement.} The author was supported by an EPSRC Advanced Fellowship.

\end{document}